\documentclass[showpacs,aps,prd,superscriptaddress,nofootinbib,floatfix,amsmath,amssymb]{revtex4}

\usepackage{graphicx}
\usepackage{dcolumn}
\usepackage{bm}

\begin{document}

\title{Rare decay $Z\to\bar\nu\nu\gamma\gamma$ via quartic gauge boson couplings}

\author{M. A. P\'erez}
\email{mperez@fis.cinvestav.mx} \affiliation{Departamento de
F\'{\i}sica, CINVESTAV, Apdo. Postal 14-740, 07000 M\'exico}
\author{G. Tavares-Velasco}
\email{gtv@itzel.ifm.umich.mx}
\affiliation{Instituto de F\'{\i}sica y Matem\'aticas, Universidad
Michoacana de San Nicolas de Hidalgo, Apdo. Postal 2-82, C.P.
58040, Morelia, Michoac\'an, M\'exico}
\author{J. J. Toscano}
\email{jtoscano@buap.fcfn.mx} \affiliation{Facultad de Ciencias
F\'\i sico Matem\' aticas, Benem\' erita Universidad Aut\' onoma
de Puebla, Apartado Postal 1152, 72000, Puebla, Pue., M\' exico}
\date{\today}

\begin{abstract}
We present a detailed calculation of the rare decay
$Z\to\bar\nu\nu\gamma\gamma$ via the quartic neutral gauge boson
coupling $ZZ\gamma\gamma$ in the framework of the effective
Lagrangian approach. The current experimental bound on this decay
mode is then used to constrain the coefficients of this coupling.
It is found that the bounds obtained in this way, of the order of
$10^{-1}$, are weaker than the ones obtained from the analysis of
triple-boson production at the CERN $e^-e^+$ collider LEP-2.
\end{abstract}


\maketitle

\section{Introduction}

One of the most sensitive probes of physics beyond the standard
model (SM) is provided by neutral gauge boson self couplings
\cite{Ellison}. In the SM, trilinear and quartic neutral gauge
boson couplings (TNGBCs and QNGBCs) $V_iV_jV_k$ and
$V_iV_jV_kV_l$, with $V_i= Z, \gamma$, vanish at tree level and
their radiative corrections are known to be rather small, of the
order of $10^{-8}$ - $10^{-10}$ \cite{Hernandez,Gounaris}.
Deviations from the SM predictions for TNGBCs and QNGBCs might
point to new interactions such as those arising from
strongly-interacting electroweak models \cite{Han}, a fourth
family of chiral fermions with SM assignments of quantum numbers
\cite{Hernandez}, and the heavy fermions arising in the minimal
supersymmetric standard model (MSSM) \cite{Gounaris}. An
interesting feature of QNGBCs involving at least one photon
field stems from the fact that they are genuine in the sense that
arise from effective operators that do not induce any trilinear gauge boson coupling.
Therefore these genuine QNGBCs must be constrained from processes
other than the ones used to constrain trilinear gauge boson couplings,
such as boson-pair fusion or triple-boson production \cite{Acciari}. This is to be
contrasted with the case of the quartic $WW\gamma\gamma$ coupling,
which can arise from operators that also induce the trilinear
$WW\gamma$ coupling, which means that any constraint on the latter
can be immediately translated into a bound on the former.
Furthermore, while the quartic neutral $ZZZZ$ coupling can be
induced at tree-level, for instance by the exchange of a heavy
scalar boson, QNGBCs involving at least one photon field can only
arise at one-loop level or higher order in any renormalizable
theory because of electromagnetic gauge invariance. In particular,
QNGBCs have been constrained from the CERN $e^+e^-$ collider LEP-2
data on triple gauge boson production \cite{Acciari}. Considerable
work has also been devoted to the analysis of QNGBCs through
different processes at the next linear $e^+e^-$ collider as well
as $\gamma\gamma$, $e\,\gamma$ and hadron colliders
\cite{Han,Eboli}.

In this note we will present a detailed calculation of the
$ZZ\gamma\gamma$ coupling contribution to the rare decay $Z \to
\bar{\nu} \nu \gamma\gamma$. The current experimental limit on this decay mode
 \cite{Adriani,Hagiwara} will then be used to
constrain the coefficients of this QNGBC. To our knowledge, this
calculation has never been presented in the literature. Our
analysis will proceed in the same line as those presented in our
previous works \cite{Larios,Larios2}, where we obtained bounds on
TNGBCs from the $Z \to \bar{\nu} \nu\gamma$ decay mode
\cite{Larios} and on neutrino-photon interactions from
$Z \to \bar{\nu} \nu\gamma\gamma$ \cite{Larios2}. We will
find that the bounds obtained in this way rely on very few
assumptions, though they are weaker than the ones obtained from
the analysis of $e^+e^-\to Z\gamma\gamma$ and $e^+e^-\to
W^+W^-\gamma$ data at LEP-2 \cite{Acciari}. The organization of
the paper is as follows. In section II we present a short
description of the effective Lagrangian for the $ZZ\gamma\gamma$
coupling. Section III is devoted to present the calculation of the
rare decay $Z\to\bar\nu\nu\gamma\gamma$. Finally, our conclusions
are presented in Section IV.

\section{The effective coupling $ZZ\gamma\gamma$}

When parametrizing physics beyond the Fermi scale in a
model-independent manner by means of the effective Lagrangian
technique there are two alternatives, i.e. the underlying new
physics can be assumed to be either of decoupled or nondecoupled
nature \cite{lagef}. In the decoupling scenario it is assumed that
the Higgs mechanism is realized in nature, thereby requiring the
existence of at least one (relatively) light Higgs boson. In this
case the low-energy theory is renormalizable {\it \`a la} Dyson,
the decoupling theorem remains valid, and the effective Lagrangian
is constructed out of operators respecting the $SU_L(2)\times
U_Y(1)$ symmetry linearly. In this scenario the virtual heavy
physics effects cannot affect dramatically the low-energy
processes: the impact of new physics might become important only in
those processes that are absent or very suppressed within the SM
\cite{lagef}. On the other hand, another possibility (the
nondecoupling scenario) arises if the Higgs boson is very heavy or
does not exist at all. There follows that the low-energy theory is
nonrenormalizable due to the absence of the Higgs boson. This
class of new physics can be assumed to be responsible for the
symmetry breaking of the electroweak sector. In this scenario the
$SU_L(2)\times U_Y(1)$ symmetry is nonlinearly
realized\cite{lagef}.

The lowest-dimension operators that induce the $ZZ\gamma \gamma$
coupling have dimension six (eight) in the nonlinear (linear)
realization. This means that any new physics effects arising from
this coupling are likely to become more evident in the nonlinear
scenario since in the linear one they are suppressed by higher
powers of the new physics scale $\Lambda$. Therefore, in this work
we will concentrate on the nonlinear scenario.  Furthermore, only
those operators that respect the custodial $SU_C(2)$ and the discrete
$C$ and $P$ symmetries will be considered. It turns out that the
operators that violate the custodial symmetry are tightly
constrained by the $\rho$ parameter. In the nonlinear scenario
there are fourteen dimension-six operators that induce the
$ZZ\gamma \gamma$ coupling at tree level \cite{Boudjema}. This was
discussed to a large extent in Ref. \cite{Boudjema} and we will
not dwell on this issue here. We rather focus on the Lorentz
structure induced for the $ZZ\gamma\gamma$ coupling. In the
unitarity gauge ($U=1$), there are only two independent Lorentz
structures for this coupling induced by dimension-six operators
\cite{Boudjema}:

\begin{equation}
\label{eflag} {\cal L}_{ZZ\gamma \gamma}=-\frac{e^2}{16\,
\Lambda^2\,c_W^2}a_0 F_{\mu \nu}F^{\mu \nu}Z^\alpha
Z_\alpha-\frac{e^2}{16\,\Lambda^2\,c_W^2}a_c F_{\mu \nu}F^{\mu
\alpha}Z^\nu Z_\alpha,
\end{equation}

\noindent where $F_{\mu \nu}=\partial_\mu A_\nu-\partial_\nu
A_\mu$ and $Z_{\mu \nu}=\partial_\mu Z_\nu-\partial_\nu Z_\mu$. We
have followed closely the notation introduced in \cite{Boudjema}.
Below we will  proceed to compute the decay $Z\to \nu \bar\nu
\gamma \gamma$ using the above effective interaction. The
experimental limit on this decay will then be used to constrain
the coefficients $a_0/\Lambda^2$ and $a_c/\Lambda^2$.

\section{The rare decay $Z\to \bar\nu\nu \gamma\gamma$}

\begin{figure}[hbt!]
\includegraphics[width=2.5in]{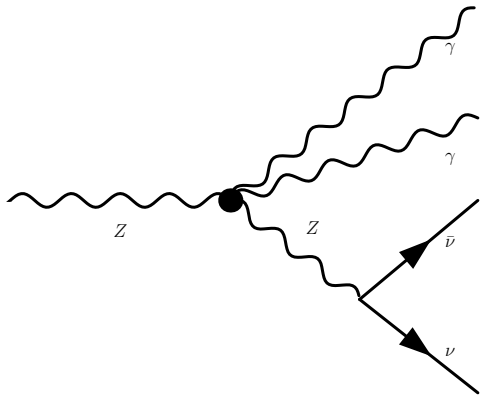}
\caption{\label{znngg}Contribution from the effective
$ZZ\gamma\gamma$ coupling to the rare $Z\to \bar\nu\nu
\gamma\gamma$ decay.}
\end{figure}

We now turn to outline the calculation of the decay width for
$Z\to \bar\nu\nu \gamma\gamma$, which is somewhat similar to the
one presented in Ref. \cite{haber} for the rare decay $Z\to \bar
\nu\nu AA$ in two Higgs doublet models, with $A$ the $CP$-odd
neutral scalar. In Ref. \cite{Larios2}, the experimental bound on
the rare $Z\to \bar\nu\nu \gamma\gamma$ decay was used to
constrain the neutrino-photon interactions $\bar\nu\nu\gamma$ and
$\bar\nu\nu\gamma\gamma$. Here we will make an analysis along the
same lines but focus on the purely bosonic $ZZ\gamma\gamma$
coupling, which contributes to the $Z\to \nu\bar\nu \gamma\gamma$
decay through the Feynman diagram shown in Fig. \ref{znngg}.

The 4-vectors of the participating particles will be denoted as
follows $Z(p)\to\bar \nu(p_1)\nu(p_2)\gamma(k_1)\gamma(k_2)$. The
Feynman rule for the effective vertex
$Z(q_\mu)Z(q^\prime_\nu)\gamma(k_\alpha)\gamma(k^\prime_\beta)$
is straightforwardly obtained from Eq. (\ref{eflag}):

\begin{eqnarray}
&&\frac{i\,e^2}{8\,\Lambda^2}
\bigg\{4\,a_0\,g_{\mu\nu}\,\left(k\cdot k^\prime\,g_{\alpha\beta}-
k^\prime_\alpha k_\beta\right) + a_c\,\Big[k\cdot
k^\prime\left(g_{\mu\alpha}g_{\nu\beta}+g_{\mu\beta}g_{\nu\alpha}\right)+
g_{\alpha\beta}\left(k_\mu\,k^\prime_\nu+k_\nu\,k^\prime_\mu\right)\nonumber\\
&&-k_\beta\left(g_{\alpha\mu}\,k^\prime_\nu+g_{\alpha\nu}\,k^\prime_\mu\right)
-k^\prime_\alpha\left(g_{\beta\mu}\,k_\nu+g_{\beta\nu}\,k_\mu\right)\Big]\bigg\}
\end{eqnarray}

\noindent where all the momenta are directed inward. The decay width can
then be written as

\begin{equation}
\Gamma(Z\to \bar\nu \nu\gamma\gamma)=\frac{1}{(2\pi)^8 2^5
m_Z}\int|\overline{\cal M}|^2 \delta^{(4)}(p-\sum_{i=1}^4
q_i)\prod_{i=1}^4\frac{d^3q_i}{2\,q_i^0},
\end{equation}

\noindent with $q_i=p_1,\,p_2,\,k_1$ and $k_2$ for $i=1,\, 2,\, 3$
and $4$, respectively.  By the usual method and after a lengthy
calculation we can obtain the squared amplitude. It reads

\begin{equation}
|\overline{\cal M}|^2=\frac{2}{3}\left(\frac{g}{4\,
c_W}\right)^2\frac{\Xi_{\mu\nu}\,p_1^\mu
p_2^\nu}{((p-k_1-k_2)^2-m_Z^2)^2}, \label{mcuad}
\end{equation}

\noindent where we have introduced the definition

\begin{equation}
\Xi_{\mu\nu}=\left(\frac{\pi\,\alpha}{m_Z\Lambda^2
c_W^2}\right)^2\left(|4\,a_0+a_c|^2(k_1\cdot
k_2)^2\,A_{\mu\nu}+2\,|a_c|^2\left(m_Z^2\,(k_1\cdot k_2)+2\,
k_1\cdot p \;k_2 \cdot p\right)B_{\mu\nu}\right),
\end{equation}

\noindent with $A_{\mu\nu}=m_Z^2\, g_{\mu \nu}+2\,p_\mu p_\nu$ and
$B_{\mu\nu}={k_1}_\mu {k_2}_\nu+{k_1}_\nu {k_2}_\mu$. In Eq.
(\ref{mcuad})  a factor of 3 is included in the denominator as we
are averaging over the $Z$ boson polarizations; also, a factor of
2 has been introduced to account for two identical particles in
the final state. The integration over $p_1$ and $p_2$ can be
carried out straightforwardly with the aid of the following result
\cite{singh}

\begin{equation}
I^{\mu\nu}=\int \frac{d^3p_1}{p_1^0}
\frac{d^3p_2}{p_2^0}\delta^{(4)}\left(Q-p_1-p_2\right)p_1^\mu
p_2^\nu=\frac{\pi}{6}\left(Q^2 g^{\mu\nu}+2\,Q^\mu Q^\nu\right),
\end{equation}

\noindent where $Q=p-p_1-p_2$. Once we are done with the
integration over $p_1$ and $p_2$, there still remains to integrate
over $k_1$ and $k_2$. To this end we will work in the center of
mass frame of the $Z$ boson. We find it useful to define the
following  variables $\xi=2\,p \cdot k_1/m_Z^2=2\,k_{1}^0/m_Z$,
$\eta=2\,p \cdot k_2/m_Z^2=2\,k_2^0/m_Z$, and
$\omega=(1-\cos\theta)/2$. The decay width can thus be expressed
as

\begin{equation}
\Gamma(Z\to
\bar\nu\nu\gamma\gamma)=\left(\frac{m_Z^2\,\alpha}{\Lambda^2\,
c_W^2}\right)^2\frac{m_Z\,\alpha}{2\,(2^8\, 3\, \pi \,c_W\,s_W
)^2}\int_\Omega h(\xi,\eta,\omega)\, d\xi\, d\eta\, d\omega,
\end{equation}

\noindent where

\begin{equation}
h(\xi,\eta,\omega)=\frac{\xi^3\,\eta^3}{(\xi\,\eta\,\omega-\xi-\eta)^2}\,\left(
|4\,a_0+a_c|^2\,f(\xi,\eta,\omega)+|a_c|^2\,g(\xi,\eta,\omega)\right),
\end{equation}

\begin{equation}
f(\xi,\eta,\omega)=\omega^2\,\left( 12 + \xi^2 - \eta\,\left( 12 -
\eta \right) - 2\,\xi\,\left( 6 - \eta - 4\,\omega\,\eta \right)
\right),
\end{equation}

\noindent and

\begin{equation}
g(\xi,\eta,\omega)=4\,\left( 1 + \omega \right) \,\left( 1 +
\omega - 2\,\omega\,\xi - 2\,\eta\,\omega\,\left( 1 - \omega\,\xi
\right)  \right).
\end{equation}

\noindent It can be shown that the integration region $\Omega$ is
given by \cite{singh}

\begin{subequations}
\begin{eqnarray}
0\leq\omega\leq 1\quad&{\mathrm{when}}&\quad 0\leq\xi\leq 1-\eta,
\\
\frac{\eta+\xi-1}{\eta\,\xi}\leq\omega\leq
1\quad&{\mathrm{when}}&\quad 1-\eta\leq\xi\leq 1,
\end{eqnarray}
\end{subequations}

\noindent together with $0\leq\eta\leq1$. After numerical
integration we are left with

\begin{equation}
\Gamma(Z\to \bar\nu \nu\gamma\gamma)=\left(\frac{1\;{\rm GeV}
}{\Lambda}\right)^4
\left(N_{0c}\,|4\,a_0+a_c|^2+N_{c}\,|a_c|^2\right)\;\;{\rm GeV},
\label{result}
\end{equation}

\noindent with $N_{0c}\approx 3.46\times 10^{-6}$ and
$N_{c}\approx 10.31\times 10^{-6}$. In addition, if lepton
universality is assumed, Eq. (\ref{result}) is to be multiplied by 3
to account for all of the known neutrino species. From the LEP-2
data, the L3 collaboration set the following limit on
$Z\to\bar\nu\nu\gamma\gamma$ \cite{Adriani}

\begin{equation}
\mathrm{BR}(Z\to\bar\nu\nu\gamma\gamma)\leq3.1 \times 10^{-6}.
\label{expbound}
\end{equation}

\begin{figure}[hbt!]
\includegraphics[width=2.5in]{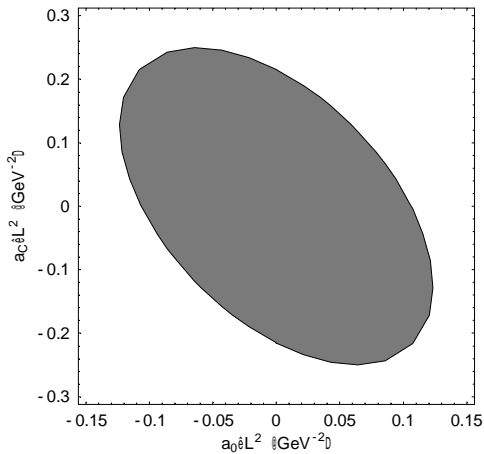}
\caption{\label{area}Allowed area (gray region) in the
$a_0/\Lambda^2$ vs. $a_c/\Lambda^2$ plane from the experimental
bound on the $Z\to \bar \nu \nu \gamma\gamma$ rare decay.}
\end{figure}

\noindent Assuming that either $a_0$ or $a_c$ is dominant we
obtain the following bounds

\begin{subequations}
\begin{eqnarray}
\frac{|a_0|}{\Lambda^2}&\leq& 0.106 \;\;{\rm GeV}^{-2}\quad{\rm if}\quad a_0\gg a_c,\\
\frac{|a_c|}{\Lambda^2}&\leq& 0.215\;\;{\rm GeV}^{-2}\quad{\rm
if}\quad a_c\gg a_0,
\end{eqnarray}
\end{subequations}

\noindent which are weaker than those obtained at LEP-2 from
$Z\gamma\gamma$ and $W^+W^-\gamma$ production \cite{Acciari}. However, it is
important to note that the bounds based on the latter processes do
not agree, as pointed out in Ref. \cite{Hagiwara}, which means
that much work along these lines is still required. In general
Eqs. (\ref{result}) and (\ref{expbound}) yield an allowed area in
the $a_0/\Lambda^2$ versus $a_c/\Lambda^2$ plane, as depicted in
Fig. \ref{area}.

We can also take a different approach and instead of bounding the
$ZZ\gamma\gamma$ coupling, we may use the most stringent bounds on
it to predict its contribution to the $Z\to\bar\nu\nu\gamma\gamma$
decay. From the most stringent experimental bound on the
$a_0/\Lambda^2$ and $a_c/\Lambda^2$ coefficients, of the order of
$10^{-2}$-$10^{-3}$ \cite{Acciari}, it follows a limit on the
contribution of the $ZZ\gamma\gamma$ coupling to the $Z\to\bar\nu\nu\gamma\gamma$ decay:

\begin{equation}
\mathrm{BR}(Z\to Z^* \gamma\gamma \to\bar
\nu\nu\gamma\gamma)\lesssim 1 \times 10^{-12}. \label{theobound}
\end{equation}

\noindent This indirect bound is above than the one found for the
contribution of the neutrino-one-photon interaction
$\bar\nu\nu\gamma$, which is of the order of $10^{-14}$
\cite{Larios2}.

\section{Concluding remarks}

In closing it is interesting to note that the $ZZ\gamma\gamma$
coupling also contributes to the rare decay $Z\to \bar l^+ l^-
\gamma\gamma$, for which an experimental bound has already been set \cite{Hagiwara}. This rare
decay mode also receives contributions from the quartic
$Z\gamma\gamma\gamma$ coupling. Therefore it would be possible, in
principle, to use this rare decay to bound such QNGBCs. As far as
the $Z\gamma\gamma\gamma$ coupling is concerned, a tighter bound
on it can be obtained from the three-body decay $Z\to
\gamma\gamma\gamma$. The latter, together with the
$e^+e^-\to\gamma\gamma\gamma$ reaction, have been studied within
the effective Lagrangian approach  \cite{Baillargeon} and we will
not repeat the same analysis here. From the result presented in
Ref. \cite{Baillargeon} a bound on the effective
$Z\gamma\gamma\gamma$ vertex can be derived, which is of the same
order of magnitude of the one obtained in this work for the $ZZ\gamma\gamma$
coupling.

Finally we would like to emphasize that the importance of studying
QNGBCs is rooted in the fact that they have a different origin
than TNGBCs, i.e. they are induced by effective operators that do
not induce any TNGBC. In this work we have found a constraint on
the $ZZ\gamma\gamma$ coupling from the $Z\to
\bar\nu\nu\gamma\gamma$ decay mode, which is weaker than the
bounds derived from the analysis of $Z\gamma\gamma$ and
$W^+W^-\gamma$ production at LEP-2. This is explained from the fact
that the decay $Z\to \bar\nu\nu\gamma\gamma$ has a suppression
factor due to the virtual $Z$ boson propagator, which suffers less
suppression when the initial $Z$ boson is allowed to be off-shell.

\begin{acknowledgments}
M.A.P. and J.J.T. acknowledge support from
Conacyt and SNI (M\'exico). The work of G.T.V. is supported by SEP-PROMEP (M\'exico).
\end{acknowledgments}

\end{document}